# Generative AI Driven Task-Oriented Adaptive Semantic Communications

Yuzhou Fu, *Student Member, IEEE*, Wenchi Cheng, *Senior Member, IEEE*, Jingqing Wang, *Member, IEEE*, Liuguo Yin, *Senior Member, IEEE*, and Wei Zhang, *Fellow, IEEE*

*Abstract*—Task-Oriented Semantic Communication (TOSC) has been regarded as a promising communication framework, serving for various Artificial Intelligence (AI) task driven applications. The existing TOSC frameworks focus on extracting the full semantic features of source data and learning low-dimensional channel inputs to transmit them within limited bandwidth resources. Although transmitting full semantic features can preserve the integrity of data meaning, this approach does not attain the performance threshold of the TOSC. In this paper, we propose a Task-oriented Adaptive Semantic Communication (TasCom) framework, which aims to effectively facilitate the execution of AI tasks by only sending task-related semantic features. In the TasCom framework, we first propose a Generative AI (GAI) architecture based Generative Joint Source-Channel Coding (G-JSCC) for efficient semantic transmission. Then, an Adaptive Coding Controller (ACC) is proposed to find the optimal coding scheme for the proposed G-JSCC, which allows the semantic features with significant contributions to the AI task to preferentially occupy limited bandwidth resources for wireless transmission. Furthermore, we propose a generative training algorithm to train the proposed TasCom for optimal performance. The simulation results show that the proposed TasCom outperforms the existing TOSC and traditional codec schemes on the object detection and instance segmentation tasks at all considered channel conditions.

## I. Introduction

With the advent of the emerging applications, such as smart monitoring, autonomous driving, and human-machine interaction, massive smart devices need to require efficient data transmission service to effectively actuate the downstream Artificial Intelligence (AI) tasks [1], thus serving users within a tolerated delay. It is envisioned that the existing wireless communications frameworks will reach their theoretical limits under limited communication resources and thus very difficult to efficiently meet stringent communication requirements, such as intelligent connection [2], high energy-efficiency [3], ultra-reliability [4], and low-latency [5]. Classic semantic communication, as an extension of Shannon theory based reliable communication, aims to convey the meaning of data to guide the receiver toward the intended goal [6]. This calls for a renewed focus on investigating semantic communication frameworks, prompted by the potential enhancements they offer to communication capability, particularly as they pertain to future sixth-generation (6G) wireless networks [7], [8].

Recently, many Deep Learning (DL) based semantic communications frameworks have been proposed to extract the semantic feature from the source data and send it to the receiver [9]–[11]. The semantic feature, which is an analog vector of lower dimensionality relative to the original source data, reserves the meaning of the source data [12]. By conveying the semantic feature, the semantic communication framework can reach higher transmission efficiency than traditional communication framework based on classic entropy coding scheme. Also, some semantic communication frameworks adopt DL based Joint Source and Channel Coding (JSCC) to learn noise resilient channel input symbols for transmitting the semantic feature over noisy channels [9], [10], thus achieving robustness to poor channel conditions. However, these previous works focus on achieving high-quality data reconstruction without establishing a strong connection to downstream AI tasks, thereby failing to serve as Task-Oriented Semantic Communication (TOSC).

Motivated by the development of AI-driven emerging applications, the research on the TOSC has become one of the hot research directions of the semantic communication [13]. Since the semantic feature of different data regions plays different roles in the correct execution of AI tasks, TOSC should extract and transmit the semantic feature to the receiver based on the task intention. There exists some literature regarding TOSC framework [14]–[17]. The authors of [14] proposed a novel semantic bit allocation model to encode semantic information with adaptive quantization, thus achieving task-oriented semantic extraction. The authors of [15] proposed a joint perception and decision scheme, which learns optimal semantic extraction scheme for classification task and aims to balance transmission latency and classification accuracy under different channel conditions. In [16], the authors proposed a semantic communication framework for image retrieval task and designed a personalized attention-based coding scheme to realize differential weight encoding of the semantic feature for important information according to user preferences. The authors of [17] proposed a Deep Reinforcement Learning (DRL) driven resource allocation scheme for TOSC to assign corresponding priority to data based on its contribution to the correct execution of AI tasks.

Yuzhou Fu, Wenchi Cheng, and Jingqing Wang are with State Key Laboratory of Integrated Services Networks, Xidian University, Xi'an, 710071, China (e-mails: fyzhouxd@stu.xidian.edu.cn; wccheng@xidian.edu.cn; jqwangxd@xidian.edu.cn).
Liuguo Yin is with Beijing National Research Center for Information Science and Technology, Tsinghua University, Beijing, 100084, China (e-mail: yinlg@tsinghua.edu.cn).
Wei Zhang is with School of Electrical Engineering and Telecommunications, the University of New South Wales, Sydney, Australia (e-mail: w.zhang@unsw.edu.au).

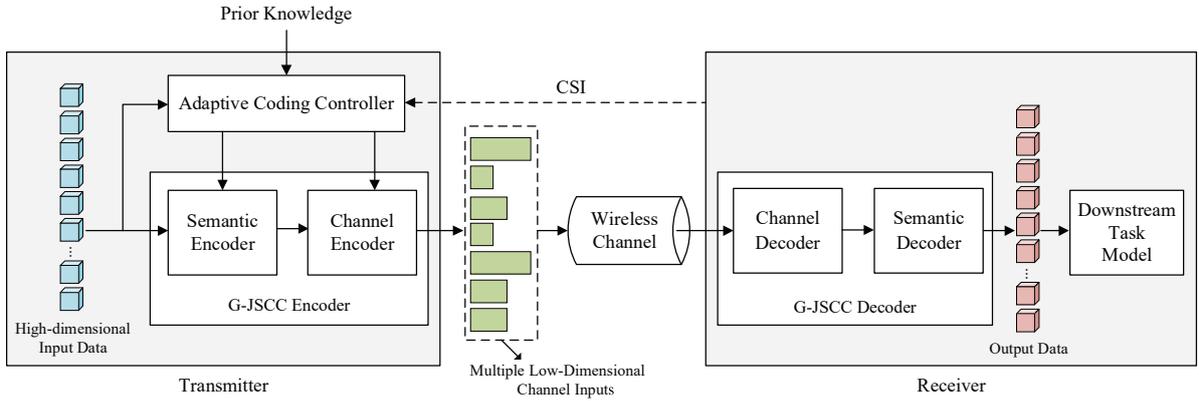

Fig. 1: The framework of our proposed task-oriented adaptive semantic communication.

Although the feasibility of TOSC has been validated, the existing works do not represent the most efficient semantic transmission schemes for serving the downstream AI tasks. The existing TOSC frameworks focus on sending the full semantic feature, which can maximize the data consistency of original data and recovered data but restrict it to reaching the performance limits of TOSC [18]. The efficient TOSC only needs to convey a subset of the full semantic feature, which can successfully guide the AI tasks toward the desired result. This subset of the full semantic feature can be called task-related semantic feature that significantly contributes to the semantic analysis of the AI tasks. Taking vehicle detection as an example for AI task, the transmission of semantic features of background and objects other than vehicles in the images is unnecessary as these features are irrelevant to the task at hand. Motivated by the objective of further investigating efficient task-oriented semantic transmission, the authors of [19], [20] focus on only extracting task-related semantic features for transmission and subsequently estimating the output data according to the received subset of full semantic features. Since the potential of semantic coding is not being fully exploited, the TOSC framework of [19] can only be applied to estimate simple content scenes, such as face and number images. Thus, it is unable to meet the requirements of various AI-driven applications in future wireless networks.

Generative AI (GAI) has significant potential in digital content generation and is expected to integrate with semantic communication to further improve its performance. However, the existing research on the GAI enhanced semantic communication aims to achieve high-quality content generation for visual experience [21] rather than serving the semantic analysis tasks. Driven by advanced GAI techniques and algorithms, such as Masked AutoEncoder (MAE) [22], Generative Adversarial Networks (GAN) [23], and Generative Pre-trained Transformer (GPT) [24], we further investigate the efficient TOSC framework, which aims to effectively facilitate the execution of AI tasks with only sending task-related semantic feature. In this paper, we propose the Task-oriented Adaptive Semantic Communication (TasCom) framework, which comprises the Generative Joint Source-Channel Coding (G-JSCC) and the Adaptive Coding Controller (ACC). In particular, the proposed G-JSCC is based on the MAE architecture to achieve efficient semantic transmission without sending full semantic feature. Then, the ACC architecture is proposed to find the optimal coding scheme to guide the semantic encoding and channel encoding based on the task requirements and channel conditions. Moreover, it enables the prioritized transmission of semantic features that significantly contribute to the AI tasks, thus efficiently utilizing limited bandwidth resources. In addition, the GPT based generative training algorithm is proposed to train our proposed TasCom neural networks to learn the optimal coded representation, thus achieving the optimal performance for the AI task.

The rest of this paper is organized as follows. Section II proposes the TasCom framework. In Section III, we introduce the design of the G-JSCC and the ACC. Then, in Section IV, the optimization objective and the generative training algorithm are proposed to train our proposed TasCom function properly. Performance analyses are given in Section V. Finally, the conclusions are drawn in Section VI.

## II. THE FRAMEWORK OF OUR PROPOSED TOASC

Figure 1 shows our proposed TasCom framework, which consists the Generative Joint Source-Channel Coding (G-JSCC) encoder, the G-JSCC decoder, and the Adaptive Coding Controller (ACC). As shown in Fig. 1, the G-JSCC encoder consists of the semantic encoder and the channel encoder for transmitting the task-related semantic feature over noisy channels. In particular, our proposed semantic encoder compresses high-dimensional input data into low-dimensional semantic features, and then only extracts the task-related semantic feature for subsequent channel encoding, thus achieving task-oriented semantic extraction. Then, the channel encoder is used to encode the task-related semantic feature into noise-robust and low-dimensional channel input with different dimensionality, thus achieving adaptive wireless transmission. Then, the ACC aims to find the optimal coding scheme to guide semantic encoder and channel encoder. The coding optimization scheme is based on the prior knowledge of downstream AI task, Channel State Information (CSI), and input data. The prior knowledge of downstream AI task includes the Neural Networks (NNs) weight of the AI task model and the delay requirement of specific AI task. We denote by input data $I$ and channel input $X$. The encoding processing of the G-JSCC encoder can be expressed as follows:

$$X = f_{\theta_2}(f_{\theta_1}(I, \gamma), \delta), \tag{1}$$

where $f_{\theta_1}(\cdot)$ represents the semantic encoder with trainable parameters $\theta_1$, $f_{\theta_2}$ represents the semantic encoder with trainable parameters $\theta_2$, the parameters of $\gamma$ and $\delta$ are produced by the ACC to be fed to the semantic encoder and the channel encoder, respectively. Then, $X$ is transmitted to the receiver over wireless channel. We adopt the Rayleigh slow fading model as the wireless channel. Let be $Y$ the received signal, which is given as follows:

$$Y = HX + G, \tag{2}$$

where $H$ and $G$ represent channel gain coefficient and the Additive White Gaussian Noise (AWGN), respectively. As shown in Fig. 1, the G-JSCC decoder consists of the semantic decoder and the channel decoder. The channel decoder is responsible for estimating the channel input from the received signal distorted by channel noise. Subsequently, the estimated values are fed to the semantic decoder to generate an output data, which can guide the AI task toward the intended results. The output data is defined as $\hat{I}$, which can be obtained by

$$\hat{I} = g_{\chi_1}(g_{\chi_2}(Y)), \tag{3}$$

where $g_{\chi_2}(\cdot)$ represents the channel decoder with trainable parameters $\chi_2$ and $g_{\chi_1}(\cdot)$ denotes the semantic decoder with trainable parameters $\chi_1$.

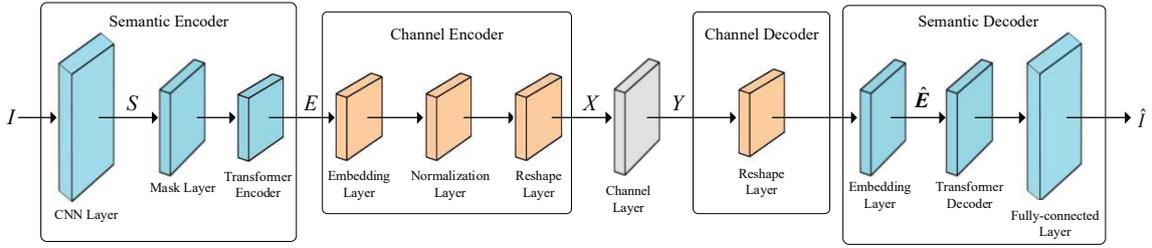

Fig. 2: The neural network architecture of the proposed G-JSCC.

### III. THE DESIGN OF THE PROPOSED TASCOM

#### A. The neural network architecture of the G-JSCC

Figure 2 shows the neural network architecture of the proposed G-JSCC. Without loss of generality, the original data is divided into non-overlapping data patches as input data, which is fed to the neural networks of the proposed G-JSCC. Thus, $I$ is made up of $M$ data patches, where $m$th ($1 \leq m \leq M$) data patch is defined as $i_m \in \mathbb{R}^l$ and $l$ is the dimensions of each data patch. The proposed semantic encoder, comprising a layer of Convolutional Neural Networks (CNNs), a mask layer, and the Transformer encoder, learns to extract task-related semantic features from the input data. Specifically, the CNNs layer learns to map input data to compact semantic feature vectors, which compress the dimensions of the input data while reserving the meaning of data. The sequence of semantic feature, denoted by $S$, which consists of $M$ semantic feature vectors. The $m$th semantic feature vector is defined as $s_m \in \mathbb{R}^q$, where $q$ represents the dimensions of each semantic feature vector and $q < l$. Then, the mask layer is used to remove useless semantic feature based on the contribution to downstream AI task, thus achieving task-oriented semantic extraction. Subsequently, the transformer encoder is used to effectively integrate the semantic information of the semantic feature vectors, thus improving the robustness to estimate data at the receiver. The task-related semantic feature is interlaced encoded to the encoded symbols. Let $E = \{e_1, ..., e_N\}$ be the sequence of encoded symbol, where $N$ is the number of encoded symbols and $N < M$. The $n$th ($1 \leq n \leq N$) encoded symbol is denoted by $e_n \in \mathbb{R}^q$. The channel encoder, comprising the embedding layer, normalization layer, and reshape layer, adopts noise-resistant coded representations of varying dimensions to transmit $E$ over noisy channels. The embedding layer is capable of mapping $E$, which has a dimensionality of $q$, to vectors with varying dimensions, thereby enabling adaptive wireless transmission. The normalization and reshape layers are combined to map the output of the embedding layer into $N$ complex channel inputs that satisfy the average transmit power constraint. The $n$th channel input is defined as $x_n \in \mathbb{R}^{k_n}$, where $k_n$ represents the dimensions of $x_n$. Then, $X$ is transmitted over the wireless channel, which is modeled as the non-trainable channel layer.

The proposed G-JSCC decoder adopts a reshape layer in the channel decoder to estimate encoded symbols based on the received complex signal. The proposed semantic decoder is made up of an embedding layer, the Transformer decoder, and a fully-connected layer. The embedding layer utilizes the learned latent representation set to expand estimated symbols, enabling the semantic decoder to generate data of the same size as the input data even without receiving full semantic features. Let $\mathcal{R} = \{r_1, ..., r_M\}$ be the latent representation set, where $r_m \in \mathbb{R}^q$ denotes the $m$th latent representation. In particular, $r_m$ is used to bridge the gap resulting from the removal of the $m$th semantic feature during semantic encoding. Thus, the embedding layer expands $N$ estimated symbols into $M$ full estimated symbols, denoted as $\hat{E} = \{\hat{e}_1, ..., \hat{e}_M\}$. Then, the Transformer decoder is utilized to decode $\hat{E}$ and generates the estimation of full semantic feature. To obtain the desired output data for the AI tasks, a fully-connected layer is employed to map the estimated full semantic feature back into data space.

#### B. The Design of the ACC architecture

To ensure the proposed G-JSCC for the optimal performance, we also propose a novel ACC architecture that de-termines the optimal coding scheme for the G-JSCC based on input data, model parameters, CSI, and tolerance delay. As shown in Fig. 3, the architecture of our proposed ACC incorporates the semantic sampling and the rate allocation parts to respectively optimize the semantic encoder and channel encoder.

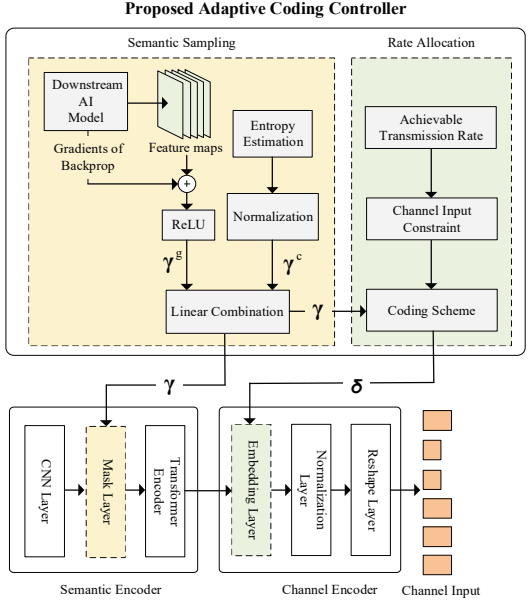

Fig. 3: The proposed ACC architecture for the semantic encoder and the channel encoder.

**Algorithm 1** A Bisection Search Method for Masking Scheme
1: **Input:** $\boldsymbol{\gamma}$, $\epsilon_{\text{th}}$, and $\boldsymbol{S}$.
2: **Initialize:** $low = 1$, $high = M$, and $\varkappa = 0$.
3: **repeat**
4:     **while** $low \leq high$ **do**
5:         $mid = \lfloor \frac{low+high}{2} \rfloor$;
6:         **if** $\gamma_{mid} > \gamma_{mid-1}$ **then**
7:             $low = mid + 1$;
8:         **else if** $\gamma_{mid} < \gamma_{mid-1}$ **then**
9:             $high = mid$;
10:         **else**
11:             $high = high - 1$;
12:         **end if**
13:         Found the minimum value $\gamma_{mid}$ within $\boldsymbol{\gamma}$;
14:     **end while**
15:     **if** $\gamma_{mid} < \epsilon_{\text{th}}$ **then**
16:         Mask layer will remove $mid$th semantic feature from $\boldsymbol{S}$;
17:         $\gamma_{mid} = 2 \times \epsilon_{\text{th}}$;
18:     **else**
19:         $\varkappa = 1$;
20:     **end if**
21: **until** $\varkappa = 1$.
22: **Output:** the task-related semantic features.

*1) Semantic Sampling Design:* The semantic sampling will generate a feature-weighted vector to guide the process of semantic encoding. The feature-weighted vector is defined as $\boldsymbol{\gamma} = (\gamma_1, ..., \gamma_M)$, where $\gamma_m \in [0, 1]$ denotes the weight value corresponding to the $m$th semantic feature. Moreover, a semantic feature with a higher weighted value indicates its greater contribution to the AI task in achieving desired results. Based on $\boldsymbol{\gamma}$, the mask layer employs a masking scheme to effectively remove useless semantic features with low weighted-values. Alternatively, we develop a bisection search approach for implementing the masking scheme, as depicted in **Algorithm 1**, where $\epsilon_{\text{th}}$ and $\varkappa$ denote the masking threshold and termination flag, respectively. In particular, the 4th to 14th lines of the proposed **Algorithm 1** implement binary search method to found the minimum value within $\boldsymbol{\gamma}$, where $\gamma_{low}$, $\gamma_{high}$, and $\gamma_{mid}$ represent the selected values within $\boldsymbol{\gamma}$ for a loop iteration, respectively. After the binary search, if the minimum value of $\boldsymbol{\gamma}$ is less than $\epsilon_{\text{th}}$, the corresponding semantic feature is removed from the semantic feature set $\boldsymbol{S}$. The current value of $\gamma_{mid}$ is set to $2 \times \epsilon_{\text{th}}$ for next loop iteration until all values of $\boldsymbol{\gamma}$ are greater than $\epsilon_{\text{th}}$.

In the semantic sampling, $\boldsymbol{\gamma}$ is a linear combination of Gradient-Weighted Mapping (GWM) and Content-Weighted Mapping (CWM). Inspired by the concept of Gradient-weighted Class Activation Mapping (Grad-CAM) [25], the model parameters can be utilized to generate GWM, which aims to predict the intention of the downstream AI task. Moreover, by employing CWM for entropy estimation, the mask layer tends to retain semantic features with high entropy, thereby enhancing the robustness of generated output data. We denote $\boldsymbol{\xi}$ as the model parameters. The process details of semantic sampling can be divided into the following steps.

**Step 1.** For a specific downstream task, the parameter-free downstream AI model is used to load $\boldsymbol{\xi}$, thus enabling the calculation of the gradient between the predicted output and the feature map of a convolutional layer. We denote $\alpha_t$ as the importance weight assigned to each activation in the $t$th feature map. The gradient values are then globally average-pooled to obtain $\alpha_t$ as follows:

$$\alpha_t = \text{GAP}\left[\frac{\partial F_{\boldsymbol{\xi}}(\boldsymbol{I})}{\partial A_t}\right], \quad (4)$$

where $F_{\boldsymbol{\xi}}(\cdot)$ represents the downstream AI task model with parameters $\boldsymbol{\xi}$, $A_t$ is the $t$th feature map in the $F_{\boldsymbol{\xi}}(\cdot)$, and $\text{GAP}[\cdot]$ represents the global-average-pooling operation. Then, we perform a weighted combination of feature maps and feed them into the Rectified Linear Unit (ReLU) layer to further highlight the semantic features that positively influence the prediction results. We define GWM as $\boldsymbol{\gamma}^{\text{g}} = (\gamma_1^{\text{g}}, ..., \gamma_K^{\text{g}})$, which can be obtained as follows:

$$\boldsymbol{\gamma}^{\text{g}} = \text{ReLU}\left(\sum_t \alpha_t A_t\right), \quad (5)$$

where $\text{ReLU}(\cdot)$ represents the ReLU operation.

**Step 2.** The CWM is obtained by applying entropy estimation to measure the information entropy of data patches. The information entropy of the $m$th data patch, denoted as $\varepsilon_m$, is defined as follows:

$$\varepsilon_m = \mathbb{E}\left[-\log p(\boldsymbol{i}_m)\right]. \quad (6)$$

We define by $\boldsymbol{\gamma}^{\text{c}} = (\gamma_1^{\text{c}}, ..., \gamma_M^{\text{c}})$ the GWM, where $\gamma_m^{\text{c}}$ represents the $m$th weighted value. Then, $\gamma_m^{\text{c}}$ is given by

$$\gamma_m^{\text{c}} = \frac{\exp(\varepsilon_m)}{\sum_{j=1}^M \exp(\varepsilon_j)}, \quad (7)$$

where $\varepsilon_j$ represents the information entropy of the $j$th ($1 \leq j \leq M$) data patch.

**Step 3.** With the liner combine of $\boldsymbol{\gamma}^g$ and $\boldsymbol{\gamma}^c$, we can obtain $\boldsymbol{\gamma}$ as follows:

$$\boldsymbol{\gamma} = \boldsymbol{\gamma}^g \mu + \boldsymbol{\gamma}^c (1 - \mu), \tag{8}$$

where $\mu \in [0, 1]$ is a tradeoff parameter.

*2) Rate Allocation Design:* The rate allocation provides a coding scheme for the channel encoder, which subsequently adjusts the dimensions of encoded symbols through the embedding layer, thus obtaining real channel input with different dimensionality to match the current channel rate. The process details of rate allocation can be divided into several steps.

**Step 1.** The rate allocation obtains the CSI to calculate the channel capacity for current transmission period. The current channel capacity is denoted as $R_{\text{cap}}$, which is given by

$$R_{\text{cap}} = B \log \left( 1 + \frac{P_T \sigma_h^2}{\sigma^2} \right) \text{ (bit/s)}, \tag{9}$$

where $B$ denotes available bandwidth, $P_T$ represents the transmission power, $\sigma^2$ is the AWGN power, and $\sigma_h^2$ is the power gain of Rayleigh fading channel. The average transmission power constraint is imposed on each channel input, ensuring that satisfies the constraint with $\frac{1}{k_n}\sum_{k=1}^{k_n}|x_k^n|^2 \leq P_T = 1$, where $x_k^n$ is the $k$th channel input symbol of $\boldsymbol{x}_n$. Considering the number of bits carried by per real channel input symbol, the achievable symbolic transmission rate in current channel capacity $R_C$ can be expressed as follows:

$$R_{\text{sym}} = \frac{B \log\left(1 + \frac{\sigma_h^2}{\sigma^2}\right)}{Q_{\text{mod}}} \text{(symbol/s)}, \tag{10}$$

where $Q_{\text{mod}}$ represents the constellation sizes of the proposed G-JSCC encoder [26].

**Step 2.** For a specific downstream AI task, the delay tolerance in the physical layer requires the receiver, which can receive all encoded symbols within a specified transmission time interval. Therefore, it is essential to determine the total number of real channel input symbols after the encoded symbols pass through the embedding layer, in terms of the given delay and current channel capacity. We denote by $D$ the delay tolerance and $L_{\max}$ the maximum-supported number of real channel input symbols. Then, $L_{\max}$ can be calculated as following:

$$L_{\max} = D \times R_{\text{sym}}. \tag{11}$$

Then, we denote by $\delta_n$ the dimensions of the $n$th real channel input symbol after the corresponding encoded symbol passed the embedding layer. In addition, $\delta_n$ should satisfy the constraint as follows:

$$\sum_{n=1}^{N} \delta_n \leq L_{\max}. \tag{12}$$

**Step 3.** Based on the $L_{\max}$ and $\boldsymbol{\gamma}$, the rate allocation adopts the coding scheme to generate an embedding token sequence, which is fed to the embedding layer to guide the channel encoding. The embedding token sequence is defined as $\boldsymbol{\delta} = (\delta_1, ..., \delta_n, ..., \delta_N)$. By leveraging linear transformation for down-sampling, the embedding layer can map a $q$-dimensional encoded symbol to a $\delta_n$-dimensional real channel input symbol according to the $\boldsymbol{\delta}$. We denote $\gamma_n \in \boldsymbol{\gamma}$ and $\gamma_t \in \boldsymbol{\gamma}$ as the weighted value of $n$th semantic feature and $t$th semantic feature, respectively. The proposed coding scheme is formulated as a min-max problem, which denoted by $\mathcal{P}$ and can be expressed as follows:

$$\mathcal{P}: \min_{\delta_n \in \boldsymbol{\delta}} \max \left( L_{\max} - \sum_{n=1}^{N} \delta_n, 0 \right), \tag{13}$$

$$\text{s.t.} \begin{cases} (\delta_n - \delta_t)(\gamma_n - \gamma_t) \geq 0, n, t \in [1, N], & (14a) \\ \delta_n \in \left\{ \frac{q}{2}, \frac{3q}{4}, q \right\}, \forall n, & (14b) \end{cases}$$

where (13) aims to optimize the utilization of available bandwidth while satisfying constraint of (14a) makes the coding scheme that tends to prioritize down-sampling the encoded symbols with low feature-weighted values until the real channel input satisfies the constraint of (12). Additionally, (14b) represents three optional dimensions of the embedding layer output. To solve $\mathcal{P}$ for the coding scheme, we can employ either a binary search algorithm or an exhaustive search algorithm to find the optimal embedding tokens $\boldsymbol{\delta}_{\text{opt}}$.

## IV. GENERATIVE TRAINING FOR PROPOSED TASCOM

### A. The optimization objective for training the TasCom

We propose a novel optimization objective to train our proposed TasCom neural networks, which can generate an output data required for the AI tasks under different $L_{\max}$ constraints. The optimization objective is the combination of three loss functions: the weighted region reconstruction loss, the feature reconstruction loss, and the GAN loss. Considering the scenarios with loose $L_{\max}$ constraint due to good channel conditions or relaxed delay requirements, we should encourage the proposed neural networks to achieve the data fidelity that refers to the output data to exactly match the bits of the input data [27], thus maximizing the performance of the AI tasks. Since different regions of the input data vary in contribution to facilitate the correct execution of AI task, we propose a weighted region reconstruction loss to compute the weighted Euclidean distance between the output data and input data. The proposed weighted region reconstruction loss, denoted by $\mathcal{L}_{\text{reg}}$, is given by

$$\mathcal{L}_{\text{reg}} = \frac{1}{M} \sum_{m=1}^{M} \gamma_m ||\boldsymbol{i}_m - \hat{\boldsymbol{i}}_m||_2^2, \tag{15}$$

where $|| \cdot ||$ is the Euclidean norm and $\hat{\boldsymbol{i}}_m$ denotes the $m$th data patch of output data. By training the proposed TasCom neural networks with $\mathcal{L}_{\text{reg}}$, it is able to maximize data fidelity, thereby enabling impeccable data reconstruction in the presence of adequate bandwidth. The drawback of data fidelity reconstruction lies in the increasing deterioration of the output data when the $L_{\max}$ constraint tightens, resulting in poor performance of downstream AI task.

Indeed, the output data exhibiting semantic feature representations similar to the input data can also facilitate downstream AI tasks toward the intended results [14], [19]. Considering the scenarios with tightened $L_{\max}$ constraint, our approach focuses on promoting similar feature representations between the output data and input data as computed by the loss network, rather than striving for an exact bits-to-bits match. Therefore, we adopt the feature reconstruction loss to train the TasCom neural networks. The feature reconstruction loss, denoted by $\mathcal{L}_{\text{fea}}$, can be expressed as follows:

$$\mathcal{L}_{\text{fea}} = ||\phi(\boldsymbol{I}) - \phi(\hat{\boldsymbol{I}})||_2^2, \tag{16}$$

where $\phi(\cdot)$ is pre-trained VGG network [28] to extract the feature representations.

In addition, we incorporate the GAN loss during the training of the TasCom neural networks, which result in the semantic consistency between input data and output data [29], even without sending full semantic features. Then, we adopt the hinge GAN loss, which denoted by $\mathcal{L}_{\text{gan}}$ and is given by

$$\mathcal{L}_{\text{gan}} = -\mathbb{E}[\log \mathcal{D}(\hat{\boldsymbol{I}}, \boldsymbol{I})], \tag{17}$$

where $\mathcal{D}$ denotes pre-trained discriminator network for the GAN loss training. To enhance stability in the training process, only the positive samples with $\mathcal{D}(\hat{\boldsymbol{I}}, \boldsymbol{I}) < 1$ and negative samples with $\mathcal{D}(\hat{\boldsymbol{I}}, \boldsymbol{I}) > -1$ will impact the gradient update of the TasCom neural networks. To train the proposed TasCom neural networks, the objective function can be formulated as the weighted combination of $\mathcal{L}_{\text{reg}}$, $\mathcal{L}_{\text{fea}}$, and $\mathcal{L}_{\text{gan}}$ as follows:

$$\min_{\theta_1, \theta_2, \chi_1, \chi_2} \left( \lambda_1 \mathcal{L}_{\text{reg}} + \lambda_2 \mathcal{L}_{\text{fea}} - \lambda_3 \mathcal{L}_{\text{gan}} \right), \tag{18}$$

where $\lambda_1$, $\lambda_2$, and $\lambda_3$ are weighing parameters.

### B. Generative Training Algorithm

We next propose a generative training algorithm in order to ensure the proposed TasCom neural networks function properly. Inspired by GPT method, the whole generative training algorithm can be divided into two-stage training. The proposed generative training first conducts stage I, which trains the G-JSCC encoder and G-JSCC decoder in an end-to-end manner while adopting random masking scheme. Thus, the G-JSCC can learn the optimal coded representations to map various input data, thus maximizing the semantic fidelity. Then, we conduct stage II to combine the pre-trained G-JSCC and the ACC architecture, and thus the whole of the TasCom neural networks are finetuned in order to fit downstream AI task models.

**Stage I Generative Pre-training for G-JSCC:** In this stage, the proposed semantic encoder randomly removes a large portion of semantic features and only operates on a small subset of the full semantic feature, laying the basis for training the optimal coded representations. We denote by $\eta$ the masking ratio, which refers to the ratio of the number of removing semantic features and full semantic features. Then, we initialize $\boldsymbol{\theta}_1$, $\boldsymbol{\theta}_2$, $\boldsymbol{\chi}_1$, $\boldsymbol{\chi}_2$, $\eta$, and $\boldsymbol{\gamma}$. Moreover, we set $\eta$ to 0.7 and $\boldsymbol{\gamma}$ is set to be random weighted values in order to achieve random masking scheme. The training algorithm is summarized in **Algorithm 2**.

The training algorithm takes place in a loop with the gradient descent, where the losses of $\mathcal{L}_{\text{reg}}$, $\mathcal{L}_{\text{fea}}$, and $\mathcal{L}_{\text{gan}}$ are minimized as (18). When the discrepancy between the input data and generated output is obtained, the parameters of $\boldsymbol{\theta}_1$, $\boldsymbol{\theta}_2$, $\boldsymbol{\chi}_1$, and $\boldsymbol{\chi}_2$ are updated by back-propagation during one training epoch. In stage I, the proposed G-JSCC is trained with learning rate $\alpha = 1e-6$, training epoch $\mathcal{E} = 200$, trained SNR $\varphi_{\text{train}} = 12$dB.

---

**Algorithm 2** Generative Pre-training for G-JSCC

1: **Input:** Dataset $\mathcal{I}$, $\mathcal{E} = 200$, $\alpha = 1e - 6$, $\eta = 0.7$, $\varphi_{\text{train}} = 12$dB, batch size $b$, $\lambda_1 = 10$, $\lambda_2 = 1$, and $\lambda_3 = 2$.
2: **Initialize:** Neural network parameters of $\boldsymbol{\theta}_1$, $\boldsymbol{\theta}_2$, $\boldsymbol{\chi}_1$, $\boldsymbol{\chi}_2$, and $\varphi_{\text{train}}$.
3: **for** training epoch= 1 to $\mathcal{E}$ **do**
4:     Sample a batch of input data $\{\boldsymbol{I}_1, ..., \boldsymbol{I}_b\} \in \mathcal{I}$;
5:     Generate $K$ random weighted value $\gamma_1, ..., \gamma_K$;
6:     To encode input data by (1);
7:     The channel input is transmitted to the receiver through noisy channel by (2);
8:     To decode received signal by (3);
9:     Obtain the generated output $\{\hat{\boldsymbol{I}}_1, ..., \hat{\boldsymbol{I}}_b\}$;
10:    Update the parameters of $\boldsymbol{\theta}_1$, $\boldsymbol{\theta}_2$, $\boldsymbol{\chi}_1$, and $\boldsymbol{\chi}_2$ by using Adam optimizer according to the proposed joint loss function (18).
11: **end for**
12: **Output:** $f_{\boldsymbol{\theta}_1}(\cdot)$, $f_{\boldsymbol{\theta}_2}(\cdot)$, $g_{\boldsymbol{\chi}_1}(\cdot)$, and $g_{\boldsymbol{\chi}_2}(\cdot)$.

---

**Stage II Training stage for the whole TasCom neural networks:** In this stage, the ACC and the G-JSCC parts are integrated with the whole TasCom neural networks. Then, a pre-trained downstream AI task model is used for providing model parameters, and thus the ACC part is evoked to produce $\boldsymbol{\gamma}$ and $\boldsymbol{\delta}$ to guide the G-JSCC encoder. With stage I training, the G-JSCC is finetuned to learn task-oriented semantic extraction as well as adaptive wireless transmission for a specific downstream AI task. The training algorithm is summarized in **Algorithm 3**, where $\mathcal{E}$ is set to 50.

---

**Algorithm 3** Finetuning for the Whole TasCom Neural Networks

1: **Input:** Dataset $\mathcal{I}$, $\mathcal{E} = 50$, $\alpha = 1e - 6$, $b$, $\lambda_{\text{MSE}} = 10$, $\lambda_{\text{FMI}} = 1$, pre-trained $f_{\boldsymbol{\theta}_1}(\cdot)$, $f_{\boldsymbol{\theta}_2}(\cdot)$, $g_{\boldsymbol{\chi}_1}(\cdot)$, and $g_{\boldsymbol{\chi}_2}(\cdot)$.
2: **for** training epoch= 1 to $\mathcal{E}$ **do**
3:     Sample a batch of input data $\{\boldsymbol{I}_1, ..., \boldsymbol{I}_b\} \in \mathcal{I}$;
4:     Input $\{\boldsymbol{I}_1, ..., \boldsymbol{I}_b\}$, $\boldsymbol{\xi}$, and CSI to the ACC;
5:     Generate $\boldsymbol{\lambda}$ and $\boldsymbol{\delta}$ by the ACC part, respectively;
6:     To encode input data by (1);
7:     The channel input is transmitted to the receiver through noisy channel by (2);
8:     To decode received signal by (3);
9:     Obtain the generated output $\{\hat{\boldsymbol{I}}_1, ..., \hat{\boldsymbol{I}}_b\}$;
10:    Update the parameters of $\boldsymbol{\theta}_1$, $\boldsymbol{\theta}_2$, $\boldsymbol{\chi}_1$, and $\boldsymbol{\chi}_2$ by using Adam optimizer according to the proposed joint loss function (18).
11: **end for**
12: **Output:** $f_{\boldsymbol{\theta}_1}(\cdot)$, $f_{\boldsymbol{\theta}_2}(\cdot)$, $g_{\boldsymbol{\chi}_1}(\cdot)$, and $g_{\boldsymbol{\chi}_2}(\cdot)$.

---

## V. NUMERICAL RESULTS

### A. Simulations Setup

In the context of automatic drive and human-machine interaction applications, the agent at the receiver aims to perceive the environment by analyzing data for AI tasks. Fig. 4 shows an application scenario example to verify the effectiveness of the proposed TasCom framework in this paper. In practice, the AI tasks always run on the edge server because the centralized database for semantic analysis is available only at the edge server [30], [31]. Hence, the proposed TasCom framework can be adopted to process the collected data and transmit them to the edge serve, thus serving the AI tasks. As shown in Fig. 4, the smart camera, serving as edge device, is equipped with the G-JSCC encoder and ACC to process the collected images and transmit them to the edge server. Then, the edge server is equipped with the G-JSCC decoder to analyze the received signal and then execute the AI tasks. Considering the IEEE 802.11 standards for the applications of low-power and low-latency, the available bandwidth $B$ is set to 20MHz [32]. The tolerance delay is set to 10ms, which serves for a wide range of surveillance applications. In the proposed G-JSCC, $Q_{\mathrm{mod}}$ is set to 8-bits constellation. Following the DL based JSCC literature [9], we refer to the ratio between the total symbols of channel input and input data (namely, $\frac{l \times M}{\sum_{n=1}^{N} k_n}$) as the Bandwidth Compression Ratio (BCR). For an image of $224 \times 224$ as input data, Table I shows the corresponding $L_{\mathrm{max}}$ constraint and achievable BCR versus channel SNR. Then, we evaluate the effectiveness of the proposed TasCom framework in terms of the performance of AI tasks such as object detection and instance segmentation. The pre-trained YOLO [33] and DETR [34] are adopted as the object detection model and the instance segmentation model to obtain the semantic label of output data, respectively. Then, we adopt the mean Average Precision (mAP) as the objective metric and also analyze the results of the object detection task and instance segmentation task as the subjective evaluation. In particular, the output data that suffers less task-related information loss could obtain a higher value of mAP.

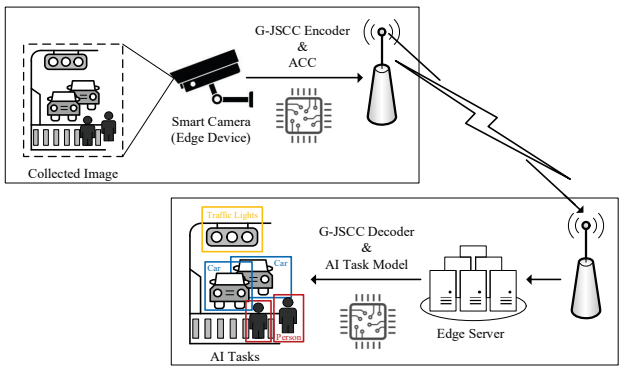

Fig. 4: An application scenario example to verify the effectiveness of our proposed TasCom framework.

### B. The Adopted Datasets

In order to train the proposed TasCom neural networks as wall as measure its performance on the AI tasks, we employ three widely used datasets:

**ImageNet-1K** [35] contains 1000 different categories in the training set and each category contains about 1000 images. The dataset is widely used for various AI-driven vision tasks, such as classification, detection, and segmentation. We use this dataset for the pre-training of our proposed G-JSCC, laying the basis for learning the optimal coded representations.

**Caltech-Pedestrians** [36] is a benchmark for pedestrian detection. It contains richly annotated video, recorded from a moving vehicle, with challenging images of low resolution and frequently occluded people. We pulled about 25000 images from the frames of the video to fine-tune and test the proposed TasCom neural networks. This dataset is appropriate for measuring the performance of TasCom framework when it is applied to the autonomous driving and smart transportation applications.

**COCO2017** [37] is a large and well annotated dataset for visual recognition, with over 330000 images and 250000 annotations. We use this dataset to fine-tune and test the proposed TasCom neural networks. Then, the dataset, which contains rich scenes, is used to measure the performance of TasCom framework when it is applied to the human-machine interaction applications.

### C. Objective Quality for Different Methods

We compare our proposed TasCom framework with the traditional codec scheme, the ideal channel scenario, the DL based ViT-ACC scheme, and the simplified TasCom scheme. Specifically, the traditional codec scheme is based on the classical separated source-channel coding, where widely used JPEG and practical LDPC are adopted as the source coding and the channel coding, respectively. The ideal channel, serving as the performance upper bound of the TasCom framework, is with unlimited channel resources, where full, noiseless, and high-dimensional semantic feature vectors can be transmitted to the receiver with full-resolution constellation. To fairly compare our proposed G-JSCC to the full semantic feature approach, we introduce ViT-ACC scheme, which follows the same structure as in Fig. 1, except that the semantic encoder and the semantic decoder are replaced by the Vision Transformer (ViT) architecture [38], thus sending full semantic feature with 4-bits constellation. To evaluate the effect of the proposed ACC architecture, we introduce simplified TasCom scheme, which adopts the coarse-grained feature maps [25] extracted from AI task model to guide the semantic encoding and the channel encoding.

Figure 5 shows the mAP achieved by various schemes versus channel SNR. For these experiments, we adopt AWGN channel and fading channel to test our proposed TasCom and other schemes, and thus the superiority of the proposed TasCom framework is fully demonstrated. It can be seen from Fig. 5 that the proposed TasCom scheme outperforms the traditional codec scheme in all channel conditions and the two datasets. In particular, the traditional codec scheme fails to reach the ideal channel baseline of semantic communication even at a channel SNR of 20dB, whereas the proposed TasCom scheme can achieve the mAP close to the ideal channel baseline when the SNR exceeds 16dB for the AWGN channel. It proves the superiority of the proposed TasCom scheme in preserving semantic fidelity in comparison to the traditional codec scheme. The simplified TasCom achieves the mAP values close to the proposed TasCom, but the increase in the performance provided by the proposed ACC scheme is visible, especially for the COCO2017 dataset. The reduced performance of the simplified TasCom may be attributed to the removal of some semantic features with key information at the semantic encoder, guided by the coarse-grained feature map. It causes the inability of the semantic decoder to estimate an output data that can achieve baseline mAP. The proposed ACC scheme introduces a certain level of content awareness, which improves the performance of the semantic decoder in estimating semantic features. Fig. 5 also shows the superiority of the proposed G-JSCC compared to the ViT-ACC with sending full semantic feature scheme. The ViT-ACC scheme fails to achieve comparable performance with the proposed TasCom across all channel SNR values in the AWGN channel, and even lags behind the traditional codec scheme at high channel SNR values. The poor performance of the ViT-ACC scheme stems from the fact that the full semantic feature transmission scheme consumes valuable bandwidth resources for transmitting task-unrelated semantic features. Also, these semantic features fail to significantly improve the performance of the AI task within limited bandwidth. In addition, the low-dimensional semantic features fail to accurately recover the meaning of original data after being seriously distorted by channel noise.

TABLE I: The $L_{\mathrm{max}}$ constraint and achievable BCR versus channel SNR.

| Channel SNR (dB) | 0 | 2 | 4 | 6 | 8 | 10 | 12 | 14 | 16 | 18 | 20 |
|---|---|---|---|---|---|---|---|---|---|---|---|
| $L_{\mathrm{max}}$ (symbol) | 0.25e+5 | 0.34e+5 | 0.45e+5 | 0.57e+5 | 0.71e+5 | 0.86e+5 | 1.01e+5 | 1.17e+5 | 1.33e+5 | 1.50e+5 | 1.66e+5 |
| BCR | 0.06 | 0.08 | 0.11 | 0.14 | 0.17 | 0.21 | 0.25 | 0.29 | 0.33 | 0.37 | 0.41 |

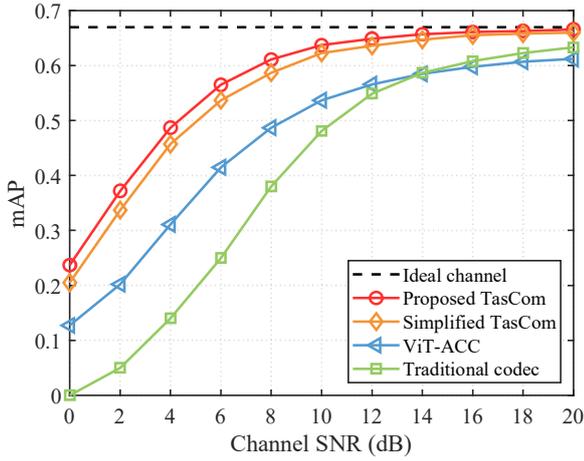
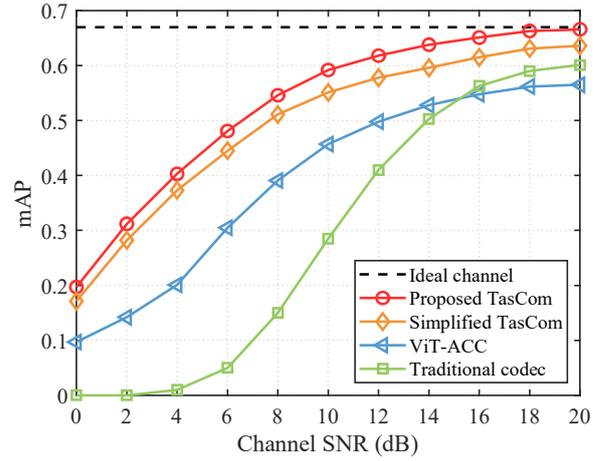
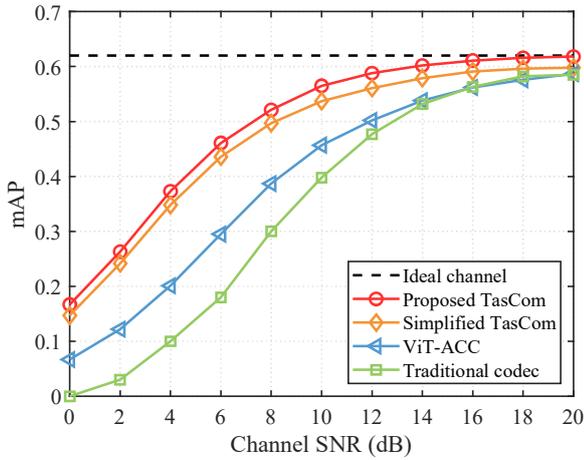
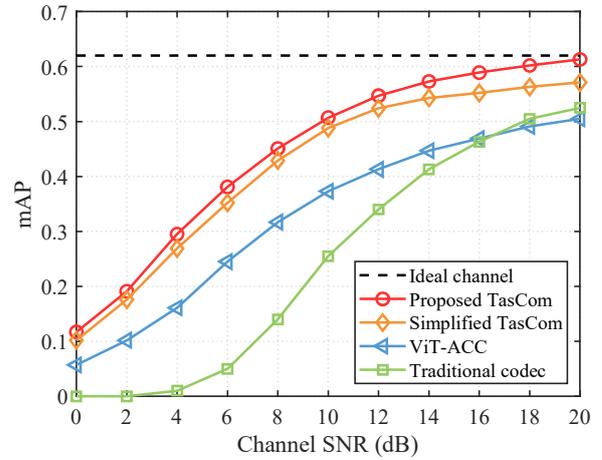

Fig. 5: Performance on the object detection task of proposed TasCom, simplified TasCom, ViT-ACC, and traditional codec schemes over AWGN and slow fading channels at the channel SNRs.

For the fading channel, the additional perturbation to the transmission symbols reduces the mAP of all schemes. However, the proposed TasCom still outperforms other schemes across all channel SNR values. The traditional codec scheme experiences cliff effect, which results in a significant reduction of the mAP when the channel SNR falls beneath the designed SNR [39]. By contrast, the mAP values achieved by the proposed TasCom achieves graceful degradation. This is because that the proposed G-JSCC demonstrates superior noise robustness. The proposed semantic encoder focuses on sending only task-related semantic features, thereby optimizing the

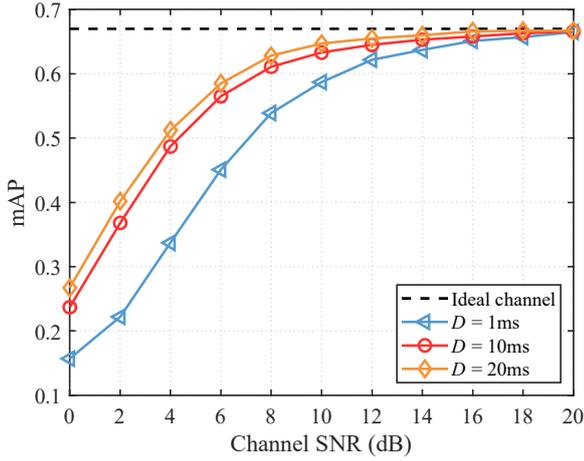 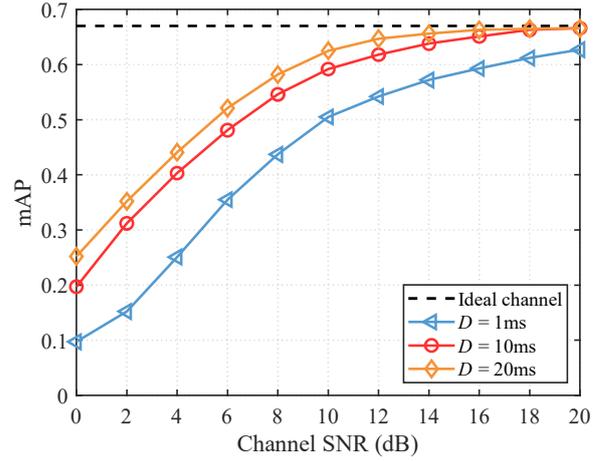

(a) Under Caltech-Pedestrians dataset and AWGN channel

(b) Under Caltech-Pedestrians dataset and fading channel

Fig. 6: Performance on the object detection task of proposed TasCom for various tolerance delays over AWGN and slow fading channels at the channel SNRs.

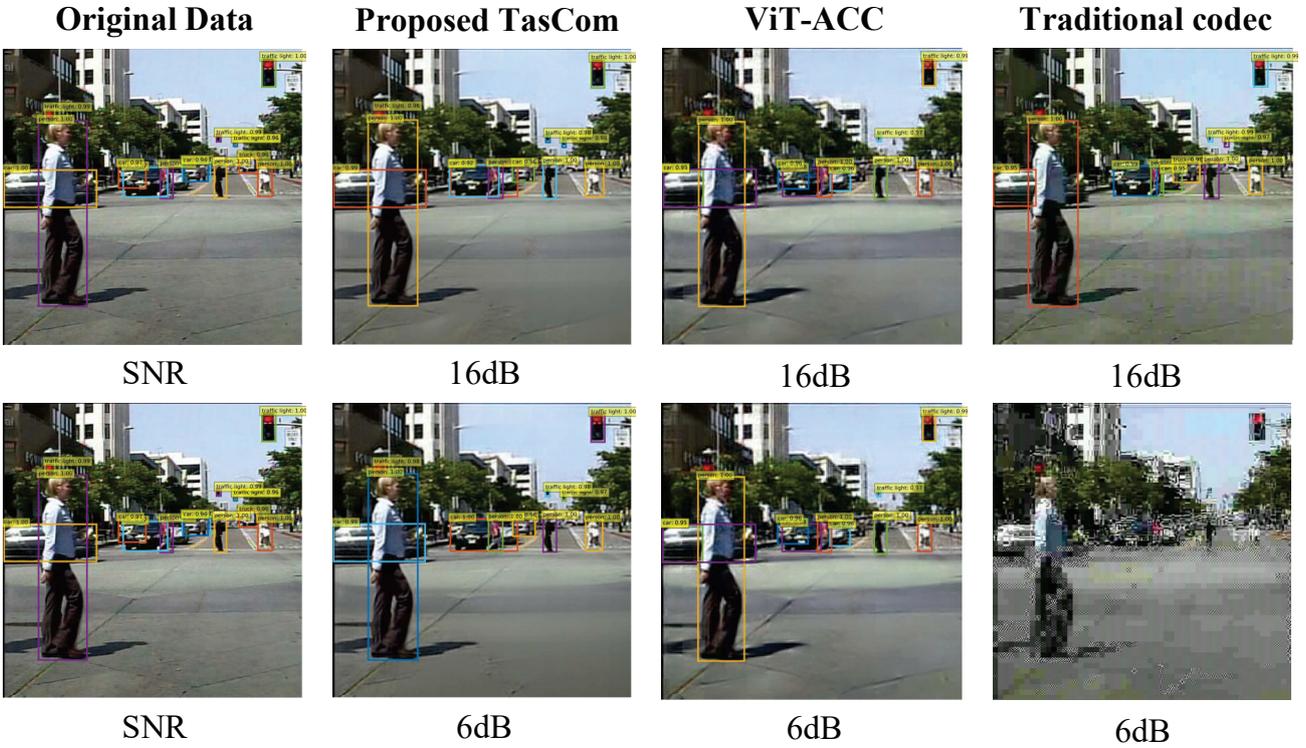

Fig. 7: Examples of proposed TasCom, ViT-ACC, and traditional codec schemes on the object detection task over fading channel for the Caltech-Pedestrians dataset.

reception of useful semantic information. Benefiting from the GAI algorithms, the proposed semantic decoder can generate the semantic feature required for the AI task without channel noise effect. The proposed TasCom can obtain the mAP values close to the ideal channel at the channel SNR of 20dB, whereas the simplified TasCom is unable to achieve the noiseless mAP values. This result further validates the effectiveness of the proposed ACC, which plays an important role in maximizing the performance of the proposed G-JSCC in the presence of fading channel.

### D. Objective Quality for Different Tolerance Delays

Then, we further investigate the impact of the tolerance delay $D$ on the mAP performance achieved by the proposed TasCom scheme, under the Caltech-Pedestrians dataset. Fig. 6 shows the mAP values achieved by proposed TasCom for various tolerance delays over AWGN and slow fading channels at the channel SNRs, where the tolerance delay is set to 1ms, 10ms, and 20ms, serving as different application requirements. The stricter tolerance delay makes the constraint of the total number of transmission symbols (12) more stringent. To com-ply with stricter tolerance delay requirements, the proposed G-JSCC aims to increase its BCR by following the guidance of the ACC scheme. It can be seen from Fig. 6 that the mAP of the proposed TasCom increases significantly with the relaxation of tolerance delay. In addition, the proposed TasCom schemes obtain the best mAP as the ideal channel for the AWGN channel at all considered tolerance delay conditions, which proves their superiority in maintaining semantic fidelity. For the fading channel, the mAP values achieved by the proposed TasCom achieve graceful improvement at all considered tolerance delay conditions. It proves the excellent robustness against the channel noise. In addition, many anti-fading techniques, such as the MIMO antenna and beamforming, can be utilized to mitigate the impact of random channel fading, thus enhancing the performance of the proposed TasCom framework under stricter tolerance delay conditions.

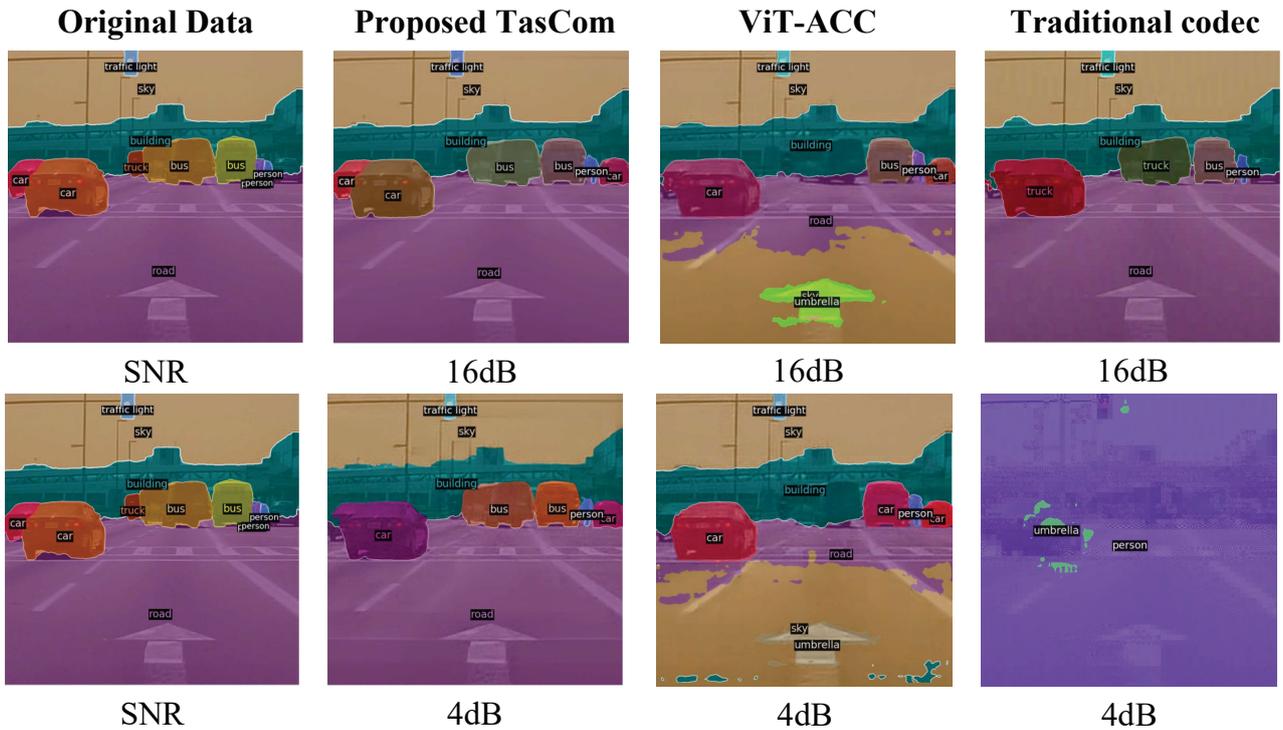

Fig. 8: Examples of proposed TasCom, ViT-ACC, and traditional codec schemes on the instance segmentation task over fading channel for the Caltech-Pedestrians dataset.

*E. Subjective Quality for Different Methods*

Figure 7 shows examples of proposed TasCom, ViT-ACC, and traditional codec schemes on the object detection task over fading channel for the Caltech-Pedestrians dataset. It can be seen from Fig. 7 that the detection result achieved by the proposed TasCom outperforms the other comparison schemes and obtains the detection result close to the original data baseline. Also, the small objects such as traffic lights and overlapped cars can be detected properly on the proposed TasCom even at the channel SNR of 6dB. However, the ViT-ACC scheme consistently fails to detect distant traffic lights and cars, exhibiting even poorer detection results compared to the traditional codec scheme at the channel SNR of 16dB. The poor detection result of ViT-ACC illustrates the limitations on effectively preserving semantic fidelity for small objects using low-dimensional feature vectors, while the traditional codec scheme based on data fidelity proves more advantageous under such circumstances. However, the traditional codec scheme fails to recover an image data to execute detection task at SNR=6dB, due to experiencing severe cliff effect.

In Fig. 8, we further show that the superiority of the proposed TasCom on the instance segmentation task compared to the ViT-ACC and traditional codec schemes. Similarly to the object detection task, the proposed TasCom still achieves the best result of the instance segmentation task at different channel SNR values. When considering the instance segmentation task required for high prediction accuracy, the ViT-ACC scheme performs worse than the traditional codec scheme at the channel SNR of 16dB. The experimental results further validate the effectiveness of task-oriented semantic transmission by the proposed G-JSCC, which becomes increasingly crucial for efficiently conveying task-related semantic information within limited bandwidth resources.

Figures 9 and 10 show the examples of proposed TasCom, ViT-ACC, and traditional codec schemes on the object detection task and instance segmentation task over fading channel for the COCO2017 dataset. To further demonstrate the superiority of the proposed GAI-driven TasCom, we employ red dotted rectangles to enclose some image regions in the original data, where the semantic features of these regions are masked in the proposed semantic encoder and thus fail to be sent to the receiver. Correspondingly, we also use the red dotted rectangles to enclose the generated image regions in the output data of the proposed TasCom. It can be observed from Fig. 9, the proposed TasCom can automatically mask the semantic features of the image regions outside the detection target, thus only sending task-related semantic feature. Furthermore, the proposed TasCom is capable of generating the image region enclosed by the red dotted rectangle without incorporating the semantic feature of its region into the semantic decoder. Even if there exist disparities between the generated region and original region, the prediction results achieved by the proposed TasCom are closet to those of the original data compared to other schemes at all channel SNR values. The outstanding detection results of the TasCom stem from its efficient task-oriented semantic transmission and proficient content genera-tion capability, thereby enhancing the performance of TOSC. In Fig. 10, both the ViT-ACC and traditional codec schemes exhibit deficiencies in terms of semantic fidelity and resilience to channel noise, regardless of whether SNR is 16dB or 6dB. We also can observe that our proposed TasCom ensures the semantic consistency between the generated data and the original data, thereby achieving identical prediction results as the original data. This again demonstrates the superiority of the proposed TasCom in comparison to the full semantic transmission and traditional codec schemes.

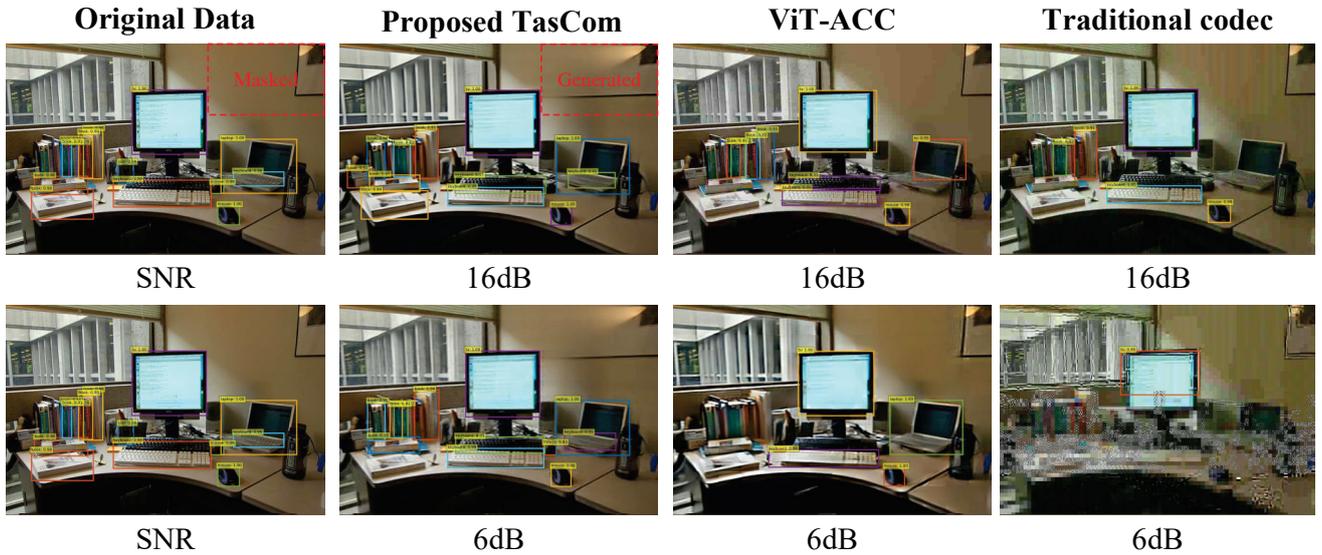

Fig. 9: Examples of proposed TasCom, ViT-ACC, and traditional codec schemes on the object detection task over fading channel for the COCO2017 dataset.

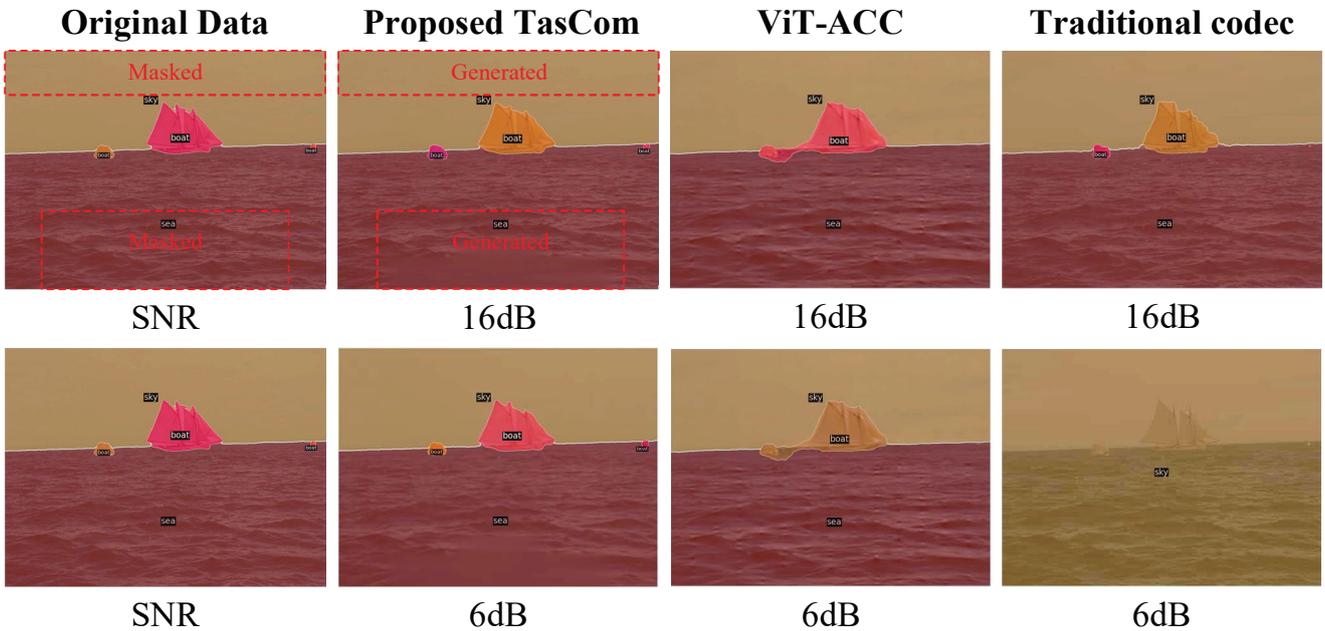

Fig. 10: Examples of proposed TasCom, ViT-ACC, and traditional codec schemes on the instance segmentation task over fading channel for the COCO2017 dataset.

## VI. Conclusion

In this paper, we have introduced the existing TOSC frameworks, which focus on sending full semantic feature but impose limitations on the performance of downstream AI tasks. To further investigate efficient TOSC, we proposed TasCom framework, which can effectively facilitate downstream AI tasks without sending full semantic feature. In particular, the proposed G-JSCC is based on the advanced MAE architecture and can generate an output data required for downstream AI tasks by only sending task-related semantic feature. We further designed the ACC architecture to optimize the coding scheme for the proposed G-JSCC in order to ensure the AI tasks toward their intended performance. Then, we proposed a novel generative training algorithm for training our proposed TasCom neural networks, which can be adapted to various AI task based applications and different noisy channels. Furthermore, we conducted an ablation study on the proposed TasCom framework and compared the alternatives under objective and subjective qualities while considering a wide range of different channel conditions. Simulation results show the superiority of the proposed TasCom framework.